\documentclass[aps,twocolumn,groupedaddress]{revtex4}
\usepackage{graphicx}
\usepackage{amsmath,amssymb}

\begin{document}

\newcommand{\etal}      {{\it et~al.}}


\title{Two different origins of the Q-slope problem in superconducting niobium film cavities for a heavy ion accelerator at CERN}



\author{A. Miyazaki$^{1,2}$ and W. Venturini Delsolaro$^{1}$}
\affiliation{$^1$CERN, Switzerland, $^2$University of Manchester, UK}


\date{\today}

\begin{abstract}
Superconducting niobium film cavities deposited on copper substrates (Nb/Cu) have suffered from strong field-dependent surface resistance, often referred to as the Q-slope problem, since their invention.
We argue that the Q-slope may not be an intrinsic problem,
but rather originates from a combination of factors which can be revealed in appropriate environmental conditions.
In this study, extrinsic effects were carefully minimized in a series of experiments on a seamless cavity.
The origin of the Q-slope in low frequency cavities is traced back to two contributions with different temperature and magnetic field dependences.
The first component of Q-slope, affecting the residual resistance, is caused by trapped magnetic flux which is normally suppressed by a magnetic shield for bulk niobium cavities.
The second, temperature dependent component of Q-slope, is similar to the medium-field Q-slope which is well known in bulk niobium cavities.
These results are compared with theoretical models and possible future studies are proposed.
\end{abstract}

\maketitle

\section{Introduction}
Superconducting RF (SRF) cavities are becoming more and more attractive as a core component of modern accelerators.
So far, the technology of choice for most realizations was based on high purity niobium sheets as the raw material for manufacturing the resonators.
Niobium film cavities deposited on copper substrates (Nb/Cu) were introduced as a promising alternative to bulk niobium cavities~\cite{benvenuti84}.
The Nb/Cu technology not only reduces production costs, but also increases thermal stability thanks to the high thermal conductivity of the film substrate.
Nb/Cu cavities were adopted for the Large Electron Positron collider (LEP-II), Acceleratore Lineare Per Ioni (ALPI), Large Hadron Collider (LHC), and High Intensity and Energy Isotope Separator On Line DEvice (HIE-ISOLDE) accelerators~\cite{kadi17}.
All these projects had relatively low field requirements.
On the other hand, the gradual increase of surface resistance at high RF field, the so-called Q-slope problem, prevents the adoption of this technology for very high-gradient accelerators, such as the International Linear Collider (ILC).
In this work, we experimentally studied Q-slopes in a Nb/Cu cavity by changing the environmental conditions.

The conventional description of the surface resistance of superconductors in the Meissner state is based on linear response theory.
At sufficiently low fields, the superconductor response can be perturbatively expanded~\cite{bardeen57}, and its first order term gives the so called BCS surface resistance~\cite{mattis58}\cite{abrikosov59}.
When the applied RF frequency is much lower than the superconducting gap energy and the operational temperature is below half of the superconducting transition temperature, 
the BCS resistance can be approximated as~\cite{halbritter70} 
\begin{equation}\label{eq:BCS-MB}
R_{\rm BCS}(T) = \frac{A_0}{T}\exp{\left(-\frac{\Delta_0}{k_{\rm B}T} \right)},
\end{equation}
where $A_0$ depends on material parameters and RF frequency, and $\Delta_0$ is the superconducting gap determined by the self-consistent gap equation of an equilibrium superconductor~\cite{bardeen57}.
A recent study by the authors on the use of this term in cavity data analysis is in Ref.~\cite{miyazaki18}.

It must be underlined that the BCS resistance, obtained by linear response theory, is independent of the applied RF field,
and, more importantly, that the theory itself is only valid at very low fields,
thus the available microscopic formulae do not tell us anything about the behavior of the surface resistance at the RF fields of interest.
The Q-slope observed in Nb/Cu cavities might be a non-linear effect beyond this microscopic description.
If the origin of the Q-slope is intrinsic and has to be explained by a microscopic theory, 
its study can be a means to examine the non-equilibrium statistical mechanics of quantum field theory,
since superconductivity is nothing but the macroscopic appearance of quantum effects in a many body system.
However, since real cavities for accelerators have huge surfaces, macroscopic inhomogeneities or thermal instabilities could also be invoked to explain the observed non-linear behavior.

This report is organized as follows.
First, the experimental setup and the main features of our cavity design are introduced.
Secondly, we report the raw results of all the measurements carried out.
Thirdly, the empirical features of the Q-slopes are exposed by a suitable data analysis.
Finally, results are discussed in terms of the historical context together with physics implications:
some theoretical models proposed previously are excluded or partially supported by our data, suggesting possible future prospects.

\section{Experimental}
\begin{figure}[h]
\includegraphics[width=40mm]{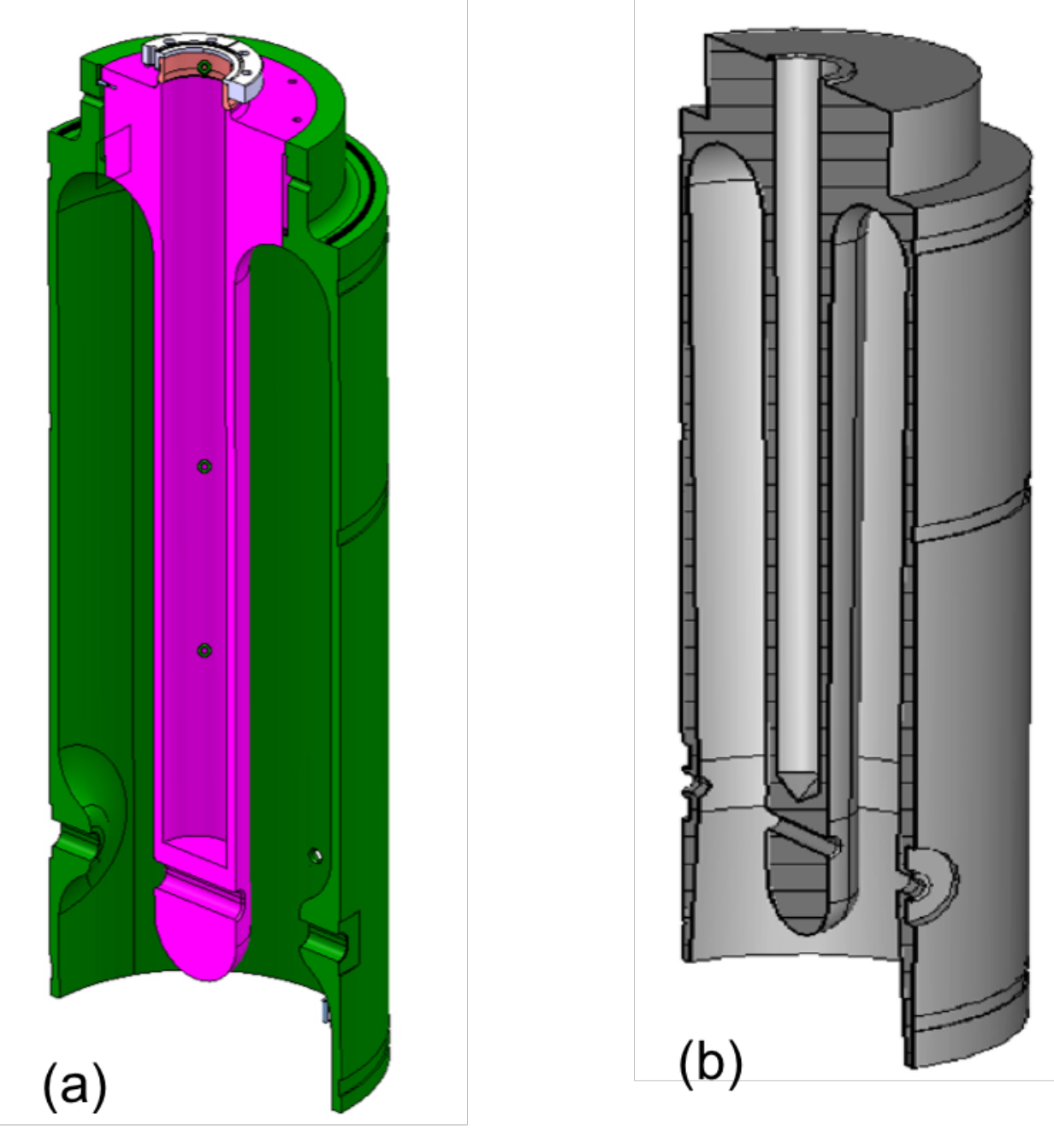}
\caption{
Cross-section view of (a) the original welded substrate and (b) the seamless substrate. 
\label{fig:QSS}}
\end{figure}

The surface resistance was studied with a newly developed Quarter-Wave Resonator (QWR)~\cite{ben-zvi83} for the HIE-ISOLDE project.
The original design of HIE-ISOLDE cavities contained a circular electron beam weld, at the top of the copper substrate, connecting inner and outer conductors as shown in Fig.~\ref{fig:QSS} (a).
As we will discuss later, this feature induced a delicate problem in cool down dynamics and made these cavities unsuitable for a systematic study of the superconducting properties of the niobium film.

In order to improve the surface quality of the copper substrate and enhance the thermal conductance to the helium bath,
we developed a new cavity~\cite{teixeira17} machined out of a solid copper billet, in which welding processes were avoided as shown in Fig.~\ref{fig:QSS} (b).
This cavity has axial symmetry apart from the beam ports, and a virtually defect-free surface in the high-field region.

The intrinsic quality factor $Q_{0}$ of the cavity was evaluated by standard methods~\cite{miyazaki18} with a Phase Lock Loop circuit.
The resonance frequency of the cavity is $101.28$~MHz and the nominal power consumption is $10$~W, corresponding to the surface resistance of $65$~n$\Omega$ at $6$~MV/m and $4.5$~K.

The accelerating gradient $E_{\rm acc}$ depends on the definition of effective length of the cavity.
In the HIE-ISOLDE project, we conventionally define
\begin{equation}
E_{\rm acc} = \frac{\int_0^{L} \left| E_{z}(z) \right|dz}{L},
\end{equation}
where $E_{\rm z}$ is the electric field component along beam axis, and $L=0.3$~m is the inner diameter of the cavity.
The Transit Time Factor (TTF) effect was not included in this definition for simplicity.
The ratio between $E_{\rm acc}$ and the peak magnetic field $B_{\rm peak}$ on the cavity surface is $9.3$~mT/(MV/m).

The measurement was carried out for different temperatures and RF fields.
First, the accelerating field was scanned from $50$~kV/m to $6$~MV/m at $4.5$~K.
Then, $Q_{0}$ was measured at a fixed field level while the cavity was cooled down to $2.4$~K.
At this point, the field was scanned at $2.4$~K.
Finally, the cavity was warmed up to $4.5$~K, and $Q_{0}$ was again measured at another fixed field.
This cycle was repeated four times, at different RF field levels.
In a second series of measurements, the same procedure was carried out after a thermal cycle above the critical temperature $T_{\rm c}=9.5$~K, with different ambient magnetic field levels when the cavity crossed $T_{\rm c}$ during the subsequent cool down.
The ambient field was controlled by external coils wound around the cryostat, and it was measured by flux-gate sensors placed around the cavity.

\section{Results}
\begin{figure}[ht]
\includegraphics[width=90mm]{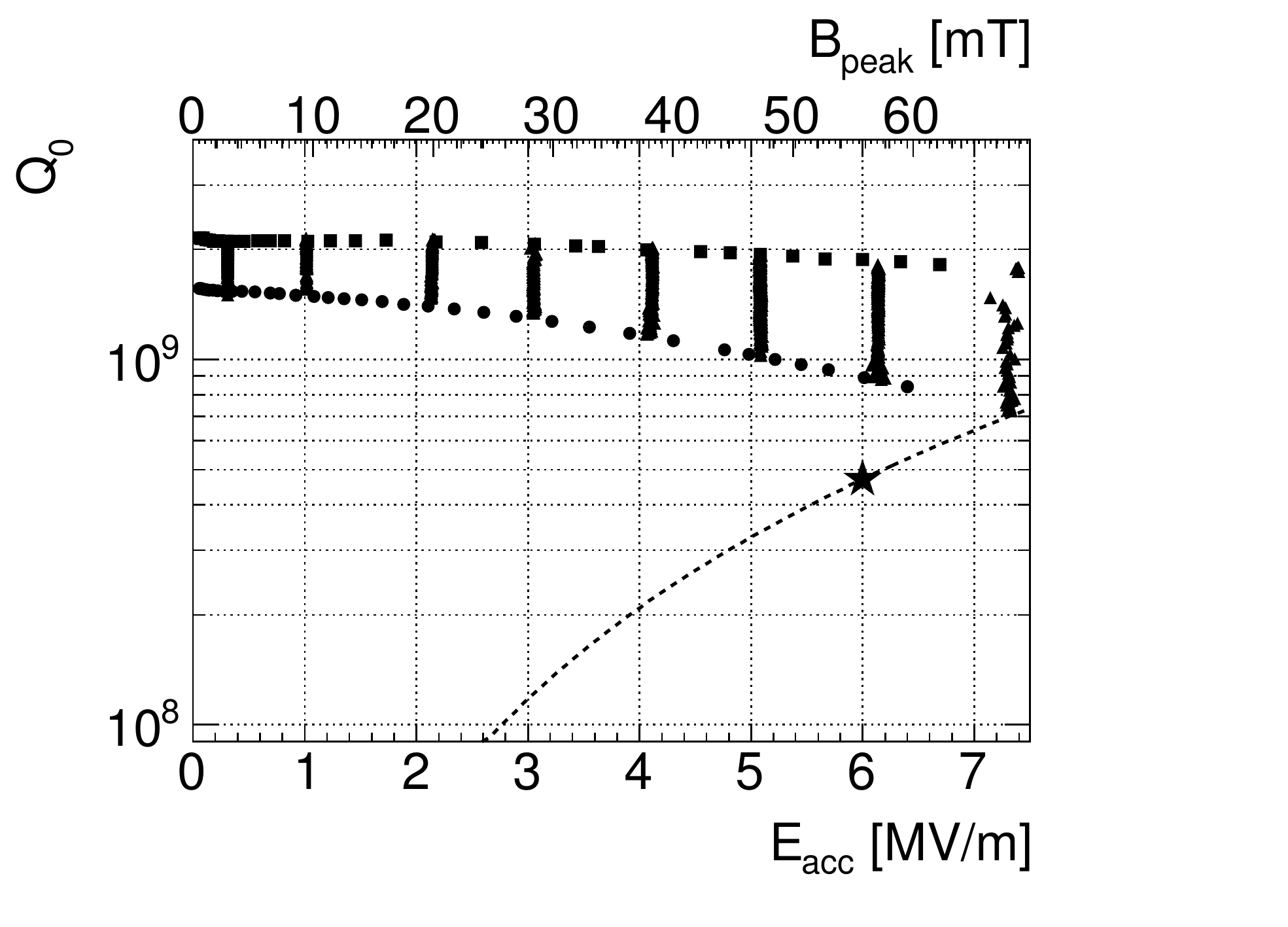}
\caption{
Quality factor as a function of accelerating gradient for a reduced external magnetic field (5~$\mu$T).
The circle points show 4.5~K and square ones show 2.4~K.
The triangle points are data of intermediate temperature at fixed fields.
The dashed line shows 10~W power consumption and the star is the nominal point.
\label{fig:Q_vs_E_reducedB}}
\includegraphics[width=90mm]{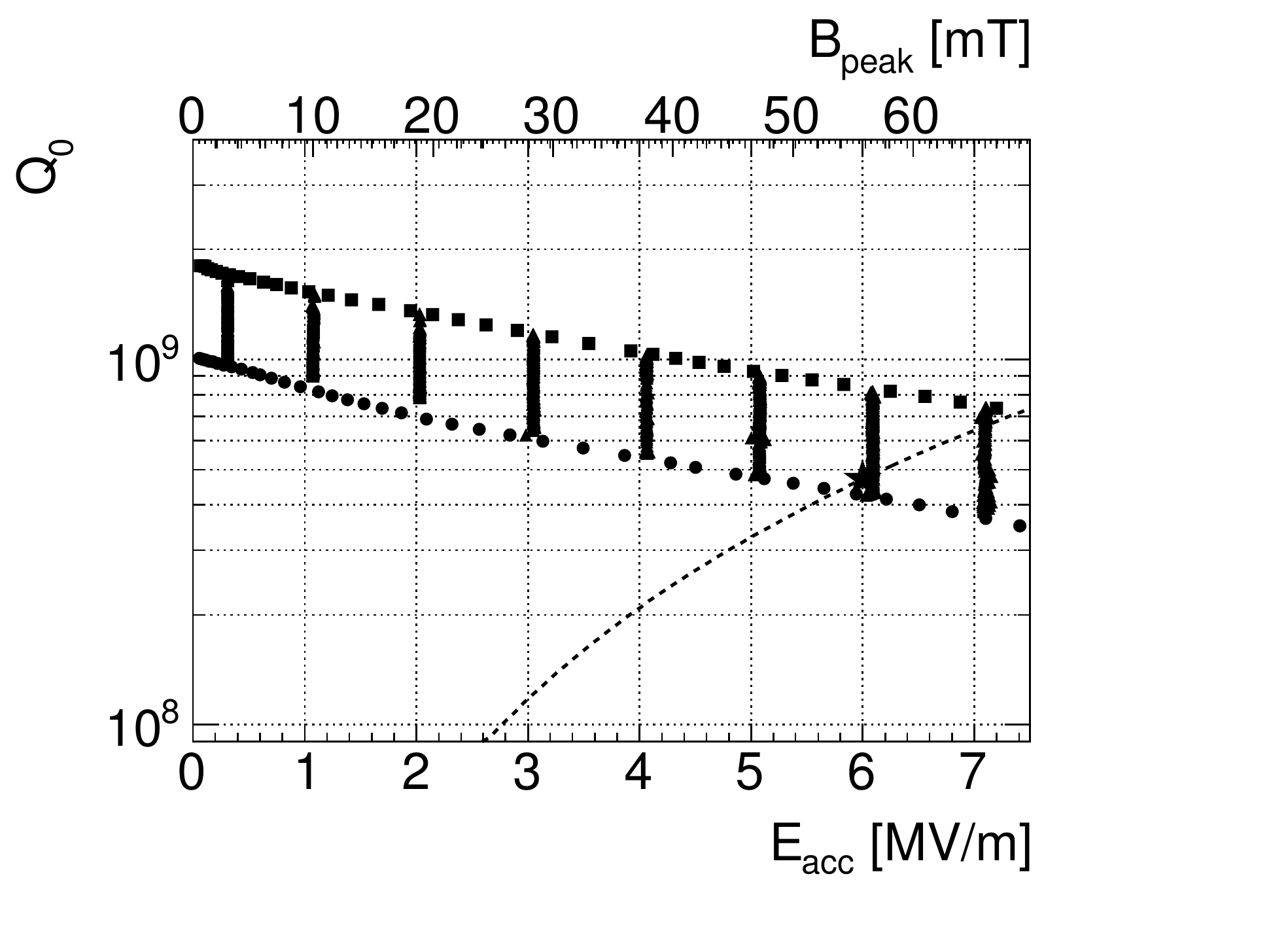}
\caption{
Quality factor as a function of accelerating gradient for an enhanced external magnetic field (100~$\mu$T).
The circle points show 4.5~K and square ones show 2.4~K.
The triangle points are data of intermediate temperature at fixed fields.
The dashed line shows 10~W power consumption and the star is the nominal point.
\label{fig:Q_vs_E_enhancedB}}
\end{figure}

Figure~\ref{fig:Q_vs_E_reducedB} shows the measured $Q_0$ as a function of $E_{\rm acc}$ for a reduced external magnetic field upon cool down through $T_{\rm c}$, and Figure~\ref{fig:Q_vs_E_enhancedB} shows the same measurements after cool down with an enhanced external field.
These results evidence that {\it a Q-slope is generated by an external field applied during cavity cooling down near $T_{\rm c}$}.

Note that the external magnetic fields were fully trapped by this cavity.
This was checked by flux-gate sensors around the cavity, showing no flux expulsion when the cavity was cooled down across $T_{\rm c}$.
In spite of the incomplete Meissner effect at the superconducting transition,
once the cavity was superconducting, and an external field was applied, the Meissner flux expulsion was clearly seen by the sensors, and no $Q_0$ degradation was observed.

The observed full field trapping during cool down may be explained by the disordered niobium film, which has presumably more pinning centers than bulk niobium cavities.
Also, the copper substrate causes a uniform temperature distribution during cool down, and the low thermal gradient does not provide enough flux repulsion force at the phase front.
Here, we do not discuss the origin of low efficiency of flux expulsion in more detail.
In the following sections, the trapped flux and the ambient fields are not distinguished.

In Fig.~\ref{fig:Q_vs_E_reducedB}, another Q-slope appears when the temperature is increased.
This effect also exists in Fig.~\ref{fig:Q_vs_E_enhancedB} but is overwhelmed by the effect of trapped flux.
The $Q_0$ measurements at a fixed field provide the temperature dependence of this Q-slope.

\section{Analysis}
Based on the raw data, detailed data analyses were carried out as follows.
The measured $Q_0$ was converted to the average surface resistance $R_{\rm s}$
\begin{equation}
\label{eq:Rs_G_Q0}
R_{\rm s} = \frac{G}{Q_0}
\end{equation}
with the geometrical factor $G$ defined by volume and surface integrals of the RF fields in the cavity and angular frequency $\omega$
\begin{equation}\label{eq:G}
G = \frac{\omega\mu_0\int_V H^2 dV}{\int_S H^2 dS}.
\end{equation}
$G$ was evaluated by commercial codes (CST MICROWAVE STUDIO~\cite{CST} and HFSS~\cite{HFSS}) and is $30.1\, \Omega$ for our cavity.
When Q-slope exists, Eq.~(\ref{eq:Rs_G_Q0}) and Eq.~(\ref{eq:G}) are not precise because of the RF field distribution over the cavity surface. 
A correction of this effect does not affect the conclusion of this analysis.
Therefore, this is omitted in the main text for simplicity.
The correction is discussed in Appendix~\ref{sec:geo-Q-slope}.

In order to separate the temperature dependent and independent terms,
the surface resistance at a given accelerating field was empirically fitted by 
\begin{equation}
\label{eq:Rs_fit}
R_{\rm s}(T) = \frac{A}{T}\exp{\left(-\frac{\Delta}{k_{\rm B}T} \right)} + R_{\rm res},
\end{equation}
with the three free parameters $A$, $\Delta$, and $R_{\rm res}$.
This function is for interpolating the data for following analysis, and does not necessarily contain any physical meanings.
However, it is worth noting that the first term of Eq.~(\ref{eq:Rs_fit}) is very similar to the linear response of the BCS theory (Eq.~(\ref{eq:BCS-MB})) in which the parameters do not depend on the RF field levels.

\subsection{Linear Q-slope in the residual resistance}\label{subsec:linear_on_RF}
\begin{figure}[ht]
\includegraphics[width=90mm]{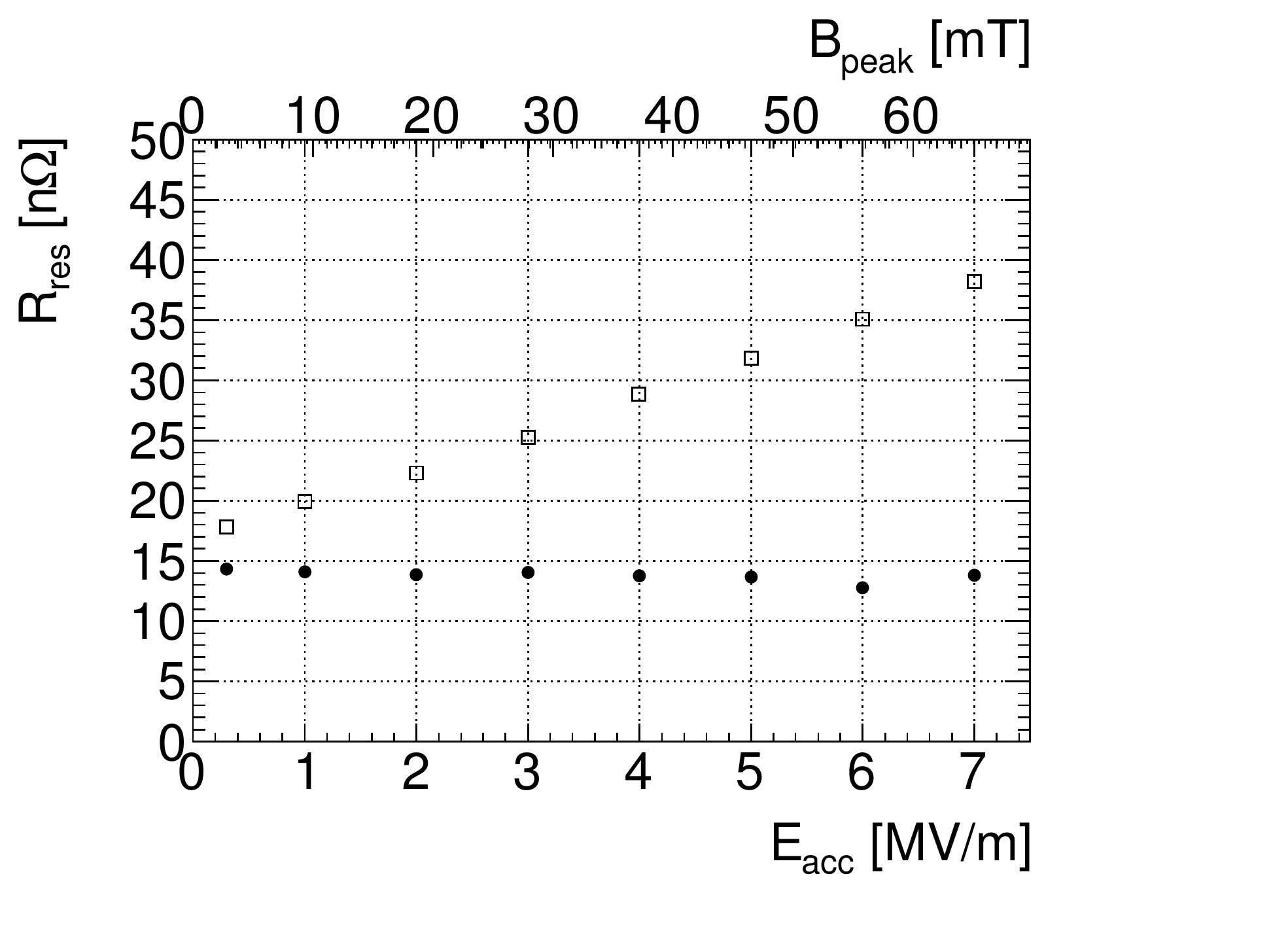}
\caption{
Residual resistance due to the external field as a function of the accelerating field.
The circles show the reduced-field (5~$\mu$T) cooling result, and blank squares show the data when the ambient field was about $100$~$\mu$T.
\label{fig:Rres_vs_Eacc}}
\end{figure}

Figure~\ref{fig:Rres_vs_Eacc} compares $R_{\rm res}$ observed with almost zero trapped flux to the case when the cavity was cooled down in $100$~$\mu$T ambient field, which was fully trapped, as discussed above.
The residual resistance caused by the trapped flux linearly depends on the RF field.
Consequently, the residual resistance $R_{\rm res}$ can be fitted by 
\begin{equation}\label{eq:def_Rs1_and_Rs0}
R_{\rm res} = R_{\rm s0} + R_{\rm s1}\times B_{\rm peak},
\end{equation}
where $R_{\rm s0}$ and $R_{\rm s1}$ are independent of the RF field.
The peak magnetic field ($B_{\rm peak}$) on the RF surface was selected as a variable.
Whenever the cavity is cooled down without shielding the ambient field, 
$R_{s1}$ dominates the total Q-slope at low temperature.

\subsection{Trapped ambient field dependence}\label{subsec:linear_on_Hext}
\begin{figure}[ht]
\includegraphics[width=90mm]{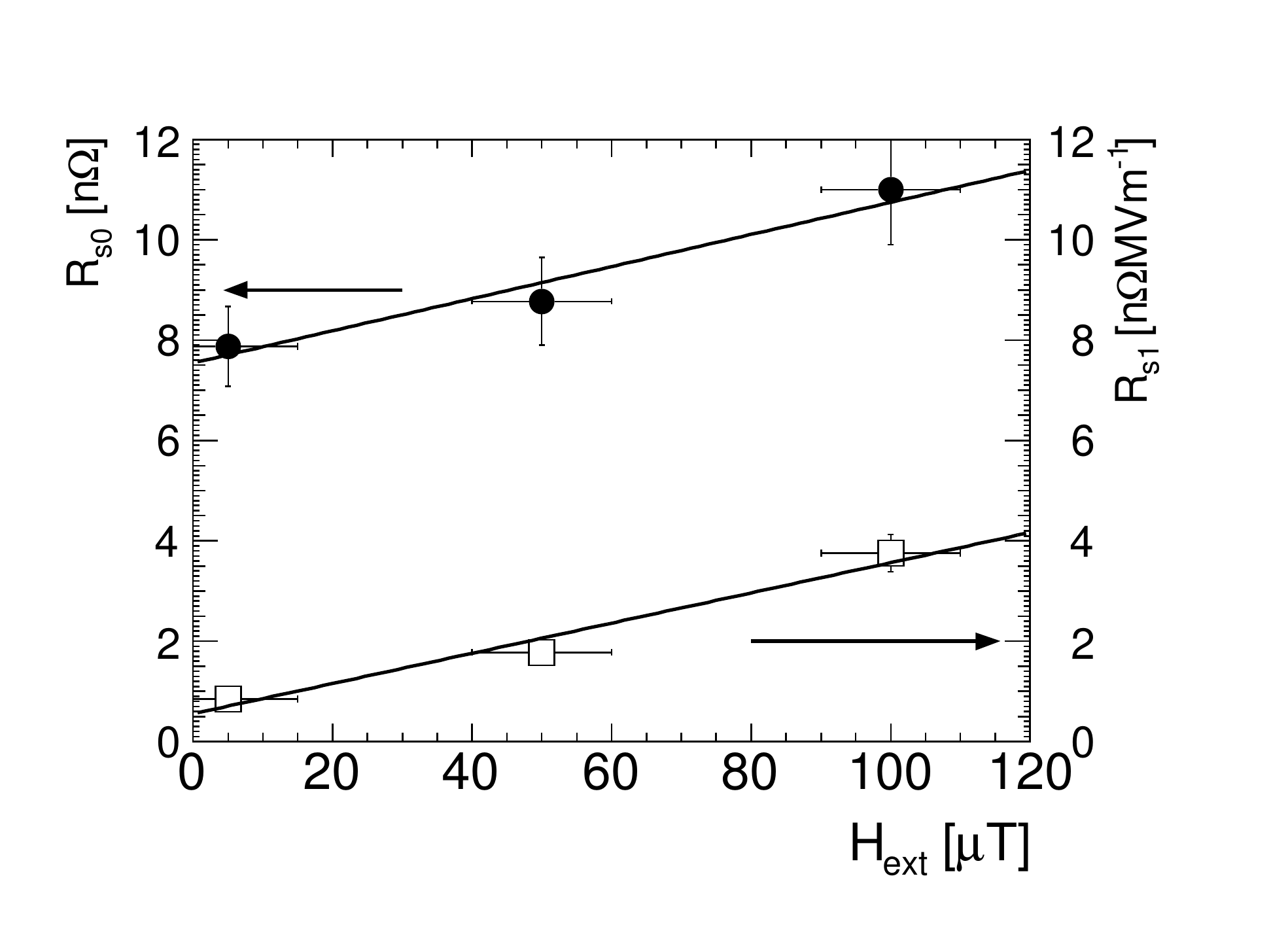}
\caption{
Residual surface resistances as a function of the trapped magnetic field.
The circles show $R_{\rm s0}$ and the squares show $R_{\rm s1}$.
\label{fig:Rs0_Rs1_vs_Hext}}
\end{figure}

$R_{\rm s0}$ and $R_{\rm s1}$ are proportional to the trapped field $H_{\rm ext}$, as shown in Fig~\ref{fig:Rs0_Rs1_vs_Hext}.
Thus, we can rewrite Eq.~(\ref{eq:def_Rs1_and_Rs0}) as
\begin{eqnarray}
R_{\rm res} & = & R_{\rm fl}(B_{\rm peak}, H_{\rm ext}) + R_{\rm res, 0}\\
            & = & \left[ R_{\rm fl, 0} + R_{\rm fl, 1} \times B_{\rm peak} \right]\times H_{\rm ext} + R_{\rm res, 0}, \label{eq:Rsfl_decomposition}
\end{eqnarray}
where the suffix ${\rm fl}$ expresses the contribution from trapped flux, and $R_{\rm res, 0}$ is the {\it intrinsic} residual resistance free from the flux effect.
In this study, no dependence of $R_{\rm res, 0}$ on the RF field was found and Q-slope in $R_{\rm res}$ was fully due to the trapped flux component $R_{\rm fl}$.
A similar function was reported in another study on 1.5~GHz elliptical Nb/Cu cavities~\cite{benvenuti99}.

\subsection{Medium field Q-slope}
\begin{figure}[ht]
\includegraphics[width=90mm]{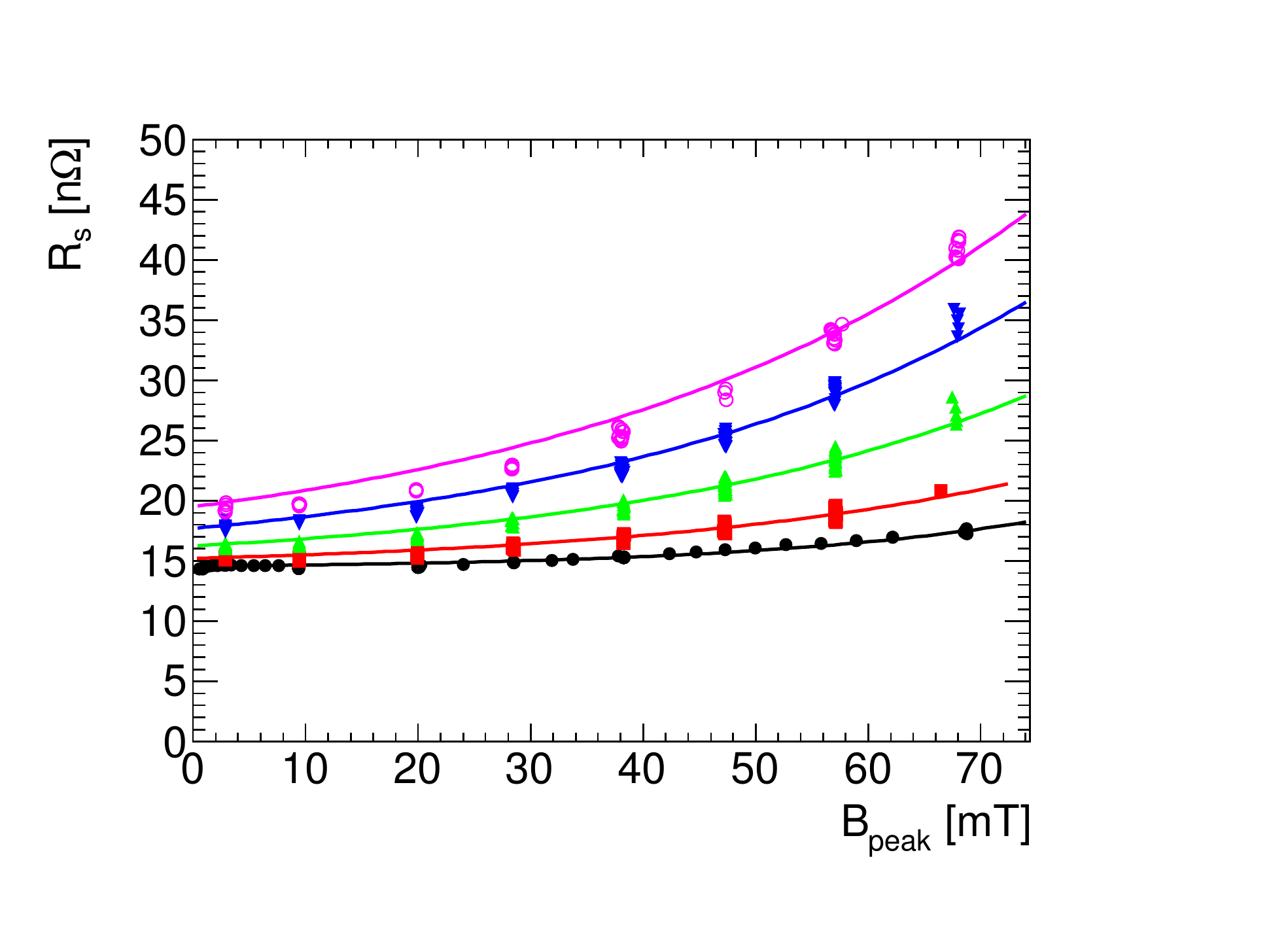}
\caption{
Temperature dependent surface resistances as a function of the RF field.
The corresponding temperatures are $2.4$, $3.0$, $3.5$, $4.0$, and $4.5$~K from the bottom.
The solid line shows the best empirical fitting at each temperature.
\label{fig:RBCS_vs_RF}}
\end{figure}

When the cavity was cooled down under well-compensated ambient field,
the linear Q-slope discussed in \ref{subsec:linear_on_RF} and \ref{subsec:linear_on_Hext} became negligibly small.
In this condition, a different type of Q-slope was exposed, which depends exponentially on the RF field as shown in Fig.~\ref{fig:RBCS_vs_RF}.
Since the temperature dependence of this component is the same as the linear response of an equilibrium BCS state, let us call this $R'_{\rm BCS}(T, B_{\rm peak})$.

The data can be fitted by an empirically found function
\begin{equation}
\label{eq:Rs_gap_reduction}
R'_{\rm BCS}(T, B_{\rm peak}) = \frac{A_0}{T}\exp{\left(\frac{\Delta_0}{k_BT} + \alpha B_{\rm peak} \right)},
\end{equation}
where $A_0$ and $\Delta_0$ are fixed at the values determined by the measurement at a low RF field, where linear response Eq.~(\ref{eq:BCS-MB}) should be still valid, and $\alpha$ is the only free parameter to be determined.

\begin{figure}[ht]
\includegraphics[width=90mm]{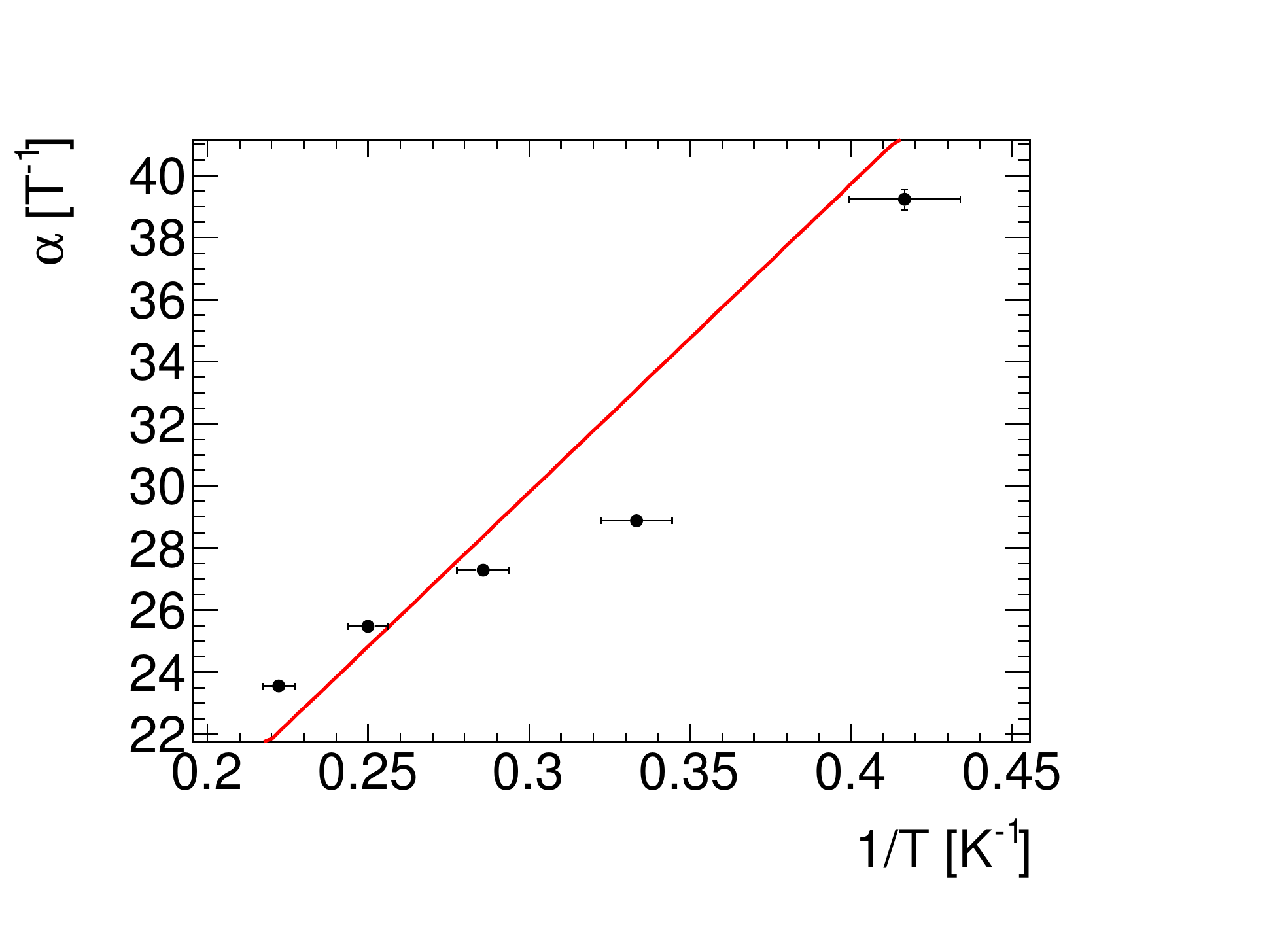}
\caption{
Temperature dependence of the empirical fitting parameter $\alpha$.
The solide line shows the best linear fit without a constant term.
\label{fig:alpha_vs_T}}
\end{figure}

The fitting parameter $\alpha$ turned out to linearly depend on the inverse of the temperature, as shown in Fig.~\ref{fig:alpha_vs_T}.
This indicates another empirical parametrization
\begin{equation}\label{eq:def_M}
\alpha = \frac{M}{k_BT},
\end{equation}
with a new constant $M$ with the dimensions of a magnetic moment.
The fitted value of $M$ is 
\begin{equation}\label{eq:value_M}
M = 1.3\,7(2)\times10^{-21}\; {\rm JT^{-1}}
\end{equation}
Thus, Eq.~(\ref{eq:Rs_gap_reduction}) can be expressed in the form of a reduced gap $\left( \Delta = \Delta_0-MB_{\rm peak} \right)$ version of Eq.~(\ref{eq:BCS-MB}).

A very similar Q-slope has been observed in bulk niobium cavities and called medium-field Q-slope~\cite{vines07}.
A conventional explanation of this Q-slope is a combination of non-linearities caused by gap reduction and thermal effects~\cite{padamsee2}.

\section{Discussion}
\subsection{Q-slope problem in Nb/Cu cavities\label{subsec:Q-slope_discuss}}
In the above, we have proposed an empirically found formula to fit the surface resistance of Nb/Cu cavities
\begin{equation}\label{eq:decomposition}
R_{\rm s}(T, B_{\rm peak}) = R'_{\rm BCS}(T, B_{\rm peak}) +  R_{\rm fl}(B_{\rm peak}, H_{\rm ext}) + R_{\rm res, 0},
\end{equation}
Each of the three components contributes about one third of the total $R_{\rm s}$ at the nominal condition of the HIE-ISOLDE cavities as summarized in Table~\ref{tab:decomposition}.
Also, two Q-slopes, $R'_{\rm BCS}(B_{\rm peak}=60)-R'_{\rm BCS}(B_{\rm peak}=0)$ and $R_{\rm fl, 1}\cdot B_{\rm peak}H_{\rm ext}$, contribute ($25$~n$\Omega$) more than half of total $R_{\rm s}$.
\begin{table}[h]
  \centering
  \begin{tabular}{lclc} \hline
    contribution      & [n$\Omega$]  &                                           &          \\ \hline
    $R'_{\rm BCS}$    & 20         & $\left\{ \begin{tabular}{@{\ }l@{}} $R_{\rm BCS}\equiv R'_{\rm BCS}(B_{\rm peak}=0)$ \\ $R'_{\rm BCS}(B_{\rm peak}=60)-R'_{\rm BCS}(B_{\rm peak}=0)$\end{tabular}\right.$   &  $\begin{tabular}{@{\ }l@{}} 5 \\ 15 \end{tabular}$    \\
    $R_{\rm fl}$     & 12          & $\left\{ \begin{tabular}{@{\ }l@{}} $R_{\rm fl, 0}H_{\rm ext}$ \\ $R_{\rm fl, 1}B_{\rm peak}H_{\rm ext}$ \end{tabular}\right.$   &  $\begin{tabular}{@{\ }l@{}} 2 \\ 10 \end{tabular}$    \\
    $R_{\rm res, 0}$  & 15         &                                           &            \\ \hline
    total             & 47         &                                           &            \\
    nominal           & 65         &                                           &            \\ \hline
  \end{tabular}
  \caption{Contribution of each component to $R_{\rm s}$ at the nominal condition of the HIE-ISOLDE cavities ($T=4.5$~K, $H_{\rm ext}=50$~$\mu$T, $B_{\rm peak}=60$~mT)}
  \label{tab:decomposition}
\end{table}

This result may shed light on the causes of Q-slope in Nb/Cu cavities at low frequency,
as it suggests that two environmental conditions $(T, H_{\rm ext})$ are the roots of the problem, which can thus be potentially observed also in bulk niobium cavities.
Indeed, some studies on bulk niobium cavities near 4.5~K show similar $R'_{\rm BCS}(T, B_{\rm peak})$~\cite{conway15}
and others at 2~K with trapped flux also show similar $R_{\rm fl}(B_{\rm peak}, H_{\rm ext})$~\cite{ciovati07}.

Normally, Nb/Cu cavities are operated at 4.5~K without magnetic shields.
Superfluid helium cooling was not considered necessary because of the good thermal conductivity of a copper substrate,
and because the BCS resistance at low field was optimized by the lower mean free path of niobium films as opposed to the bulk.
Magnetic shields are also omitted for coated cavities, because sensitivities to the trapped flux were reported to be small in the literature~\cite{benvenuti99}.
However, while the constant term of sensitivity $R_{\rm fl, 0}$ is indeed one or two orders of magnitude smaller than that of bulk niobium,
the linear term $R_{\rm fl, 1}B_{\rm peak}$ is not negligible especially when the intrinsic resistance $R_{\rm res, 0}$ is made smaller by an increased quality of the niobium film.

When the external field was well compensated and the cavity was cooled down by superfluid helium, two Q-slopes were drastically surpressed.
Although detailed field scan was done only up to $B_{\rm peak}=70$~mT as shown in Fig.~\ref{fig:Q_vs_E_reducedB},
the cavity reached a maximum $B_{\rm peak}=120$~mT.
This is the highest field value ever achieved in a Nb/Cu cavity~\cite{delsolaro18}.

One complication is cool down dynamics in the bimetal structure of Nb/Cu cavities.
The clear decomposition of $R_{\rm s}$ proposed above was only possible with the seamless QWR shown in Fig.~\ref{fig:QSS} (b) in which the temperature distribution during cooling down was made very uniform by the excellent thermal conductance between the RF surface and the helium bath.
In the welded design (Fig.~\ref{fig:QSS} (a)),  the linear component $R_{\rm s1}$ strongly depends on the cool down dynamics because the thermal conductance is limited by the welded section.
For welded cavities, $R_{\rm s1}$ could not be eliminated even by the best possible cooling down in our cryogenic system, 
and the $R'_{\rm BCS}$ component of those cavities remained masked by the effect of the cool down dynamics.

The effect of cool down dynamics in bulk niobium cavities has been intensively studied~\cite{posen16}.
Higher thermal gradients are preferred to more efficiently expel the magnetic field upon transition to the superconducting state.
On the contrary, $R_{\rm s0}$ and $R_{\rm s1}$ of Nb/Cu cavities turned out to be improved by {\it lower} thermal gradients.
This was reported by Ref.~\cite{zhang15} and recent studies will be summarized in Appendix~\ref{sec:Q-slope_welded}.
This behavior is a clear difference between Nb/Cu and bulk niobium, and implies different mechanism on flux-induced surface resistance.
One plausible hypothesis is that thermoelectric currents confined in the interface of Nb/Cu structure are trapped by the niobium film at $T_{\rm c}$.
Since the Seebeck effect is proportional to the thermal gradient, 
this hypothesis qualitatively explains the observation.
A similar phenomenon was reported in cavities where the RF surface is a layer of Nb$_3$Sn grown on niobium~\cite{posen17}.
A direct measurement of such thermoelectric current, however, could not be obtained so far.

It is worth noting that no low-field Q-slope was observed in this cavity.
In a good niobium cavity, surface resistance is usually increased when the RF field is decreased to very low fields.
The relatively high residual resistance of our sputtered cavity may prevent the observation of such phenomenon.
This is an indirect evidence that the problem of this cavity is not the medium-field Q-slope and flux-induced loss, which commonly exist even in bulk niobium cavities, but the intrinsic residual resistance without Q-slope.
Thus, further optimization of the coating parameters should be carried out to reduce the remaining normal conducting defects in the present recipe.

\subsection{Physics interpretation on the trapped flux induced Q-slope\label{subsec:trapped_flux}}
The conventional phenomenological models of the flux oscillation do not predict a linear dependence $R_{\rm fl, 1}$ on the RF field.
The models studied so far are linearized, and their  solution, the oscillation amplitude, is also proportional to the applied RF field.
In such models, the power consumption by the flux oscillation is proportional to the square of the RF field; thus, the surface resistance is a constant of the field.

The first model was proposed by Bardeen and Stephen in Ref.~\cite{bardeen65}, where flux motion driven by Lorentz force is described by effective inertial mass, viscosity, and tension.
Gittleman and Rosenblum~\cite{gittleman65} studied trapped fields above $H_{\rm c1}$ applied on thin films (PbIn and NbTa alloys of about 12.7 $\mu$m thick).
They neglected the tension term of the Bardeen-Stephen model.
Checchin {\it et al.}~\cite{checchin17} extended this model for pinned flux below $H_{\rm c1}$ in bulk niobium.
They approximated the trapped flux by a flexible tube and the tension term was still omitted.
In both cases, the oscillation amplitude of the flux is assumed to be small so that the pinning potential was approximated as harmonic.
The approach of Gurevich and Ciovati~\cite{gurevich13} took into account the tension term, and treated pinning as the Dirichlet boundary condition in which the flux line is fixed at the pinning centers.
Still, this model translates into a linear partial differential equation and the predicted surface resistance is independent of the RF field.
Heating at a hot spot caused by the trapped flux was also considered, but the resulting thermal-runaway function does not fit the linearly increasing surface resistance as a function of the RF field.

A straightforward extension of these models is to introduce non-linear terms in the equation of motion.
There are several possibilities to do this, for example
\begin{enumerate}
\item Collective weak pinning~\cite{liarte18}
\item Non-harmonic pinning potential~\cite{vaglio18}
\item Velocity dependence in the effective viscosity
\end{enumerate}
Some encouraging theoretical results have been obtained.
For example, Ref.~\cite{vaglio18} predicts positive correlation between $R_{\rm fl, 1}$ and $R_{\rm fl, 0}$.
See Appendix~\ref{sec:Q-slope_welded} for more discussion.
Parameter dependence of the models should be compared with the experimental results.
These models have different predictions on behaviors against RF frequency and mean free path, for example.
Direct measurement of the pinned flux may also provide some hints on the non-linearity of the phenomenon~\cite{embon15}.

\subsection{Physics interpretation of medium-field Q-slope}
Great care should be taken when attempting physics interpretations of $R'_{\rm BCS}$,
although our empirical fitting evokes gap reduction within the BCS framework.
Surface resistance is a macroscopic observable averaged over a cavity surface, penetration depth, and RF cycles.
Hence, the measured $R'_{\rm BCS}$ is not necessarily a mere microscopic response of the superconductor reacting against the applied RF field.
It could arise as well from macroscopic inhomogeneities.

\subsubsection{Possible extension of the BCS resistance\label{sec:gap_reduction_theory}}
If we ignore the extrinsic complications mentioned above,
a possible empirical interpretation of $R'_{\rm BCS}$ may be gap reduction within the linear response theory.

Depression of superconducting gap by a DC applied magnetic field or current was studied theoretically and experimentally~\cite{douglass61}\cite{bardeen62}\cite{meservey64}.
A gap equation modified for a current-carrying state results in an effectively reduced $\Delta$.
Tunneling experiments verified the depression of $\Delta$.
This phenomenon is more substantial when the sample is thin compared with the penetration depth and when the temperature is close to $T_{\rm c}$.
Besides the gap reduction, the density of states was found to be smeared by a DC current both in dirty and clean limits~\cite{fulde65}.

A phenomenological model of gap reduction by the RF current was proposed by Kulik and Palmieri in Ref.~\cite{kulik98}\cite{palmieri05}, where $\Delta$ is given by
\begin{equation}
\Delta = \Delta_0 - p_{\rm F}v_{\rm s},
\end{equation}
with $p_{\rm F}$ the Fermi momentum and $v_{\rm s}\propto B_{\rm peak}$ the super-current velocity.
They substituted this modified $\Delta(B_{\rm peak})$ into Eq.~(\ref{eq:BCS-MB}).
Gurevich recalculated the integral of surface resistance in momentum space followed by averaging over the penetration depth and the RF period~\cite{gurevich06} rather than using the naive substitution.
Its result is parabolic in the RF field strength at low field and is exponential at higher RF field.
In this model, a characteristic magnetic moment $M$ is 
\begin{equation}
M = \frac{\pi}{2^{3/2}}\frac{\Delta}{B_{\rm c}}\sim1.13\times10^{-22}\; {\rm JT^{-1}}.
\end{equation}
This is one order of magnitude smaller than the experimental value Eq.~(\ref{eq:value_M}).
This discrepancy has been already reported for bulk niobium, and the literature combined thermal feedback model with this non-linear BCS resistance to fit the medium-field Q-slope~\cite{vines07}.

More recently, it was suggested that the gap reduction due to the RF field may be associated with smearing of the density of states~\cite{xiao13}\cite{gurevich14}.
This phenomenon had been already shown in the DC case in Ref.~\cite{kulik98} but was not explicitly used in the derivation of surface resistance.
In this context, surprisingly, the gap reduction does exist but is overwhelmed by smearing of the density of states, which reduces the number of effective quasiparticles contributing to the RF loss, and results in a depression of surface resistance i.e. anti-Q-slope.
These models were developed to explain the anti-Q-slope observed in N-doped cavities~\cite{grassellino13}.
However, their derivation is so general that the gap reduction proposed to explain the normal medium-field Q-slope should be also hidden by the effect of density of states smearing.
Note that the formula in Ref.~\cite{gurevich14} is only valid in the dirty limit, because the Usadel equation~\cite{kopnin01} is used, but the smearing of the density of states seems a more general phenomenon regardless of the purity of the material as shown in Ref.~\cite{fulde65}.

These approaches are based on an extension of the gap reduction observed in the DC case to RF frequencies.
Another approach, valid at higher frequency but low field, was proposed by Eliashberg.
This leads to the {\it enhancement} of $\Delta$ when a high frequency field is applied~\cite{eliashberg70}.
The quasiparticle distribution function becomes non-equilibrium by RF photon absorption and effectively $\Delta$ can be larger.
Such enhancement was observed experimentally at 10~GHz through tunneling measurements where the current value is very low~\cite{kommers77}.

The above review illustrates how a field dependence of the effective superconducting gap arises in models going beyond the linear response of the equilibrium BCS state.
However, the predictive power of these procedures remains limited.
Therefore, we do not conclude that Eq.~(\ref{eq:Rs_gap_reduction}) can theoretically explain the medium-field Q-slope but just report its exponential RF field dependence and universality in low frequency Nb/Cu cavities and bulk niobium cavities. 

\subsubsection{Macroscopic models\label{subsec:V-P}}
Phenomenologically, there are several possible ways to explain $R'_{\rm BCS}(T, B_{\rm peak})$.
Here, we review thermal instability models.

Since the linear response Eq.~(\ref{eq:BCS-MB}), valid at low RF field, depends on temperature exponentially, 
Joule heating can cause thermal runaway and can eventually trigger a thermal quench in bulk niobium cavities.
However, such thermal runaway does not fit medium-field Q-slope, in which the surface resistance gradually increases from low RF fields.
The literature~\cite{vines07} combines thermal feedback with the non-linear microscopic theory.
As discussed in Sec.~\ref{sec:gap_reduction_theory}, non-linear extensions of the microscopic theory are still an open field of research.

In Nb/Cu cavities, thermal instabilities should be strongly suppressed by the good thermal conductivity of copper substrates.
Global heating was indeed excluded by a dedicated experiment measuring the helium gas pressure in an elliptical Nb/Cu cavity~\cite{junginger15}.
However, heating may still occur {\it locally}, and a small area in the normal conducting state would have a large influence on the surface resistance,
without affecting the average temperature of the cavity surface.

Recently, a new model was proposed, that takes into account defects at the interface between a niobium film and copper substrate~\cite{palmieri16}.
Using a distribution function $f(R_{\rm B})$ of thermal resistance $R_{\rm B}$ at such defects and local surface resistance $R_{\rm s}(T_0, E_{\rm acc}, R_{\rm B})$, which can be quenched by thermal feedback, the observable surface resistance is given by 
\begin{equation}
\label{eq:V-P}
\overline{R_{\rm s}}(T_0, E_{\rm acc}) = \int_{0}^{\infty} R_{\rm s}(T_0, E_{\rm acc}, R_{\rm B}) f(R_{\rm B}) dR_{\rm B},
\end{equation}
where $T_0$ is the bath temperature and $R_{\rm B}$ is a local thermal resistance which locally prevents the film from cooling down.
In this model, interface defects with bad thermal resistivity cause local quenches which do not propagate to the rest of the film.
The existence of such locally quenched defects is not excluded by the temperature measurement discussed above.

\begin{figure}[h]
\includegraphics[width=80mm]{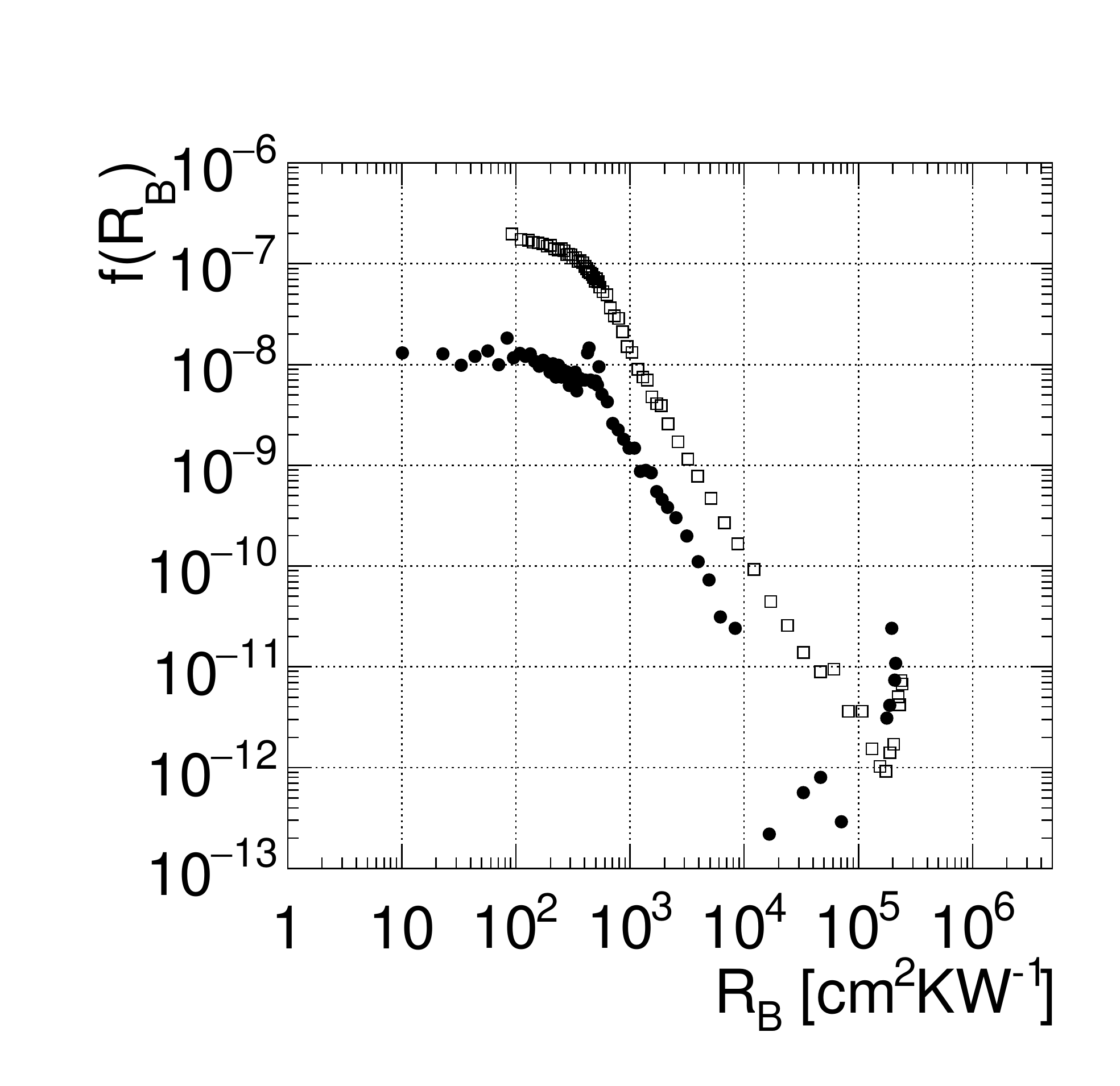}
\caption{
Converted distribution of thermal boundary resistance. The circle points show 2.4~K data and blank square ones show 4.5~K data.
\label{fig:V-P-converted}}
\end{figure}

The welded series production of the HIE-ISOLDE QWR was studied using this model~\cite{calatroni16}.
A simple algorithm was proposed to inversely solve Eq.~(\ref{eq:V-P}), and $f(R_{\rm B})$ can be reconstructed from the observed $\overline{R_{\rm s}}(T_0, E_{\rm acc})$.
We applied this formalism to the Q-slope of the seamless cavity discussed in this paper, and compared two different bath temperatures. 
Figure~\ref{fig:V-P-converted} shows that calculated $f(R_{\rm B})$s differ by one order of magnitude between 2.4~K and 4.5~K.
This is unreasonable because $f(R_{\rm B})$ is dominated by the mechanical adherence of the interface between niobium film and copper substrate,
and should not strongly depend on bath temperature.

Previous studies reported that $f(R_{\rm B})$ was independent of the bath temperature~\cite{aull16}\cite{vaglio17},
and concluded that this was an evidence of the thermal boundary problem in Nb/Cu cavities.
The analysis of one of the authors was quoted in these studies.
However, these results were without the decomposition of trapped flux effect discussed in Sec.~\ref{subsec:Q-slope_discuss}.
Before the removal of this effect, the linear component $R_{\rm fl, 1}$ dominates the Q-slope.
Since $R_{\rm fl, 1}$ does not strongly depend on temperature, the numerical inversion of Eq.~(\ref{eq:V-P}) yields $f(R_{\rm B})$ looking independent of temperature on a logarithmic scale.

There are other indirect arguments which may disfavor this model.
Thermal simulations showed that defects of small $R_{\rm B}$, which are abundant in Fig.~\ref{fig:V-P-converted}, 
are not stable above $T_{\rm c}$~\cite{Furci17}.
Also, as emphasized above, a very similar medium-field Q-slope has been observed in bulk niobium cavities; thus, a model based particularly on Nb/Cu interface may not be reasonable for this Q-slope.


\subsection{Future prospects of Nb/Cu cavities}
This study indicated that the Q-slope problem at low frequency (100~MHz) can be avoided if the copper substrate is properly designed, external magnetic fields are well shielded, and bath temperature is low enough, as for bulk niobium cavities.
This leads to promising applications of the Nb/Cu technology to low-$\beta$ accelerators for heavy ions or upstream sections of proton Linacs.
For instance, the HIE-ISOLDE post-accelerator is successfully providing radioactive beams for nuclear physics.

The application to higher frequency cavities is still uncertain.
Historically, the Nb/Cu technology has focussed on exploring various different coating methods or optimizing deposition parameters.
This study revealed the overwhelming importance of the copper substrate and supported the choice of a seamless cavity machined from a bulk copper billet.
One possibility is to make machined elliptical substrates of higher frequency.

The good thermal conduction of the seamless QWR also provided uniform temperature distribution {\it during sputtering}.
This may have some impact on the deposited film quality as well.
Clearly, the thermoelectric current and cool-down-dynamics issues deserve more dedicated studies.

\section{Conclusion}
The Q-slope of a low-frequency Nb/Cu QWR was experimentally decomposed into two phenomena.
One is responsible for a residual resistance linearly dependent on RF fields, and is caused by the trapped flux.
The theory of flux induced loss should be revised because the conventional linearized theories do not contain such RF dependence.
If the ambient field is properly shielded, this cavity, whose substrate is machined from bulk copper, is free from this Q-slope.
The other component is medium-field Q-slope which exponentially depends on both RF field and temperature.
In superfluid helium, this Q-slope is as small as in bulk niobium cavities,
and this cavity reached the world best peak magnetic field in the Nb/Cu technology.
As a similar Q-slope was also found in other cavities made of bulk niobium, this component seems a universal property, and not a unique problem of the Nb/Cu technology.
An empirical fit suggests reduction of $\Delta_0$, although a full theoretical justification is still missing.
Thermal instability models seem not satisfactory.
More dedicated studies on medium field Q-slope in more controlled environment and cavities are necessary.

\section*{Acknowledgement}
We gratefully acknowledge the contribution of our colleagues A.~Sublet, S.~Teixeira, T.~Mikkola, M.~Therasse and M.~A.~Fraser for their support in cavity design, preparation and testing. 
Our special thanks go to R.~Vaglio, M.~Arzeo, A.~E.~Ivanov and K.~Turaj for useful discussion.
We warmly thank all the technical staff at CERN for their invaluable help.

\appendix
\section{Geometrical corrections}\label{sec:geo-Q-slope}
If the surface resistance varies on the RF surface, the definition of observable surface resistance $\langle R_{\rm s} \rangle$ is
\begin{equation}
\langle R_{\rm s} \rangle = \frac{\int_S R_{\rm s}(x, y, z)H^2(x, y, z) dS}{\int_S H^2(x, y, z) dS},
\end{equation}
where $R_{\rm s}$ is the local surface resistance at the given position $(x, y, z)$ and the integral is over the RF surface.
If the material is uniform over the surface, the surface resistance does not explicitly depend on position but only through RF field $H$, i.e.
\begin{equation}
R_{\rm s}(x, y, z) = \left(R_{\rm s}\circ H\right) (x, y, z).
\end{equation}
As D.~Longuevergne showed~\cite{longuevergne13}\cite{longuevergne18}, one can change the integral variables from $(x, y, z)$ to $H$ with a distribution function $S(H)$ over the surface
\begin{equation}
\label{eq:david}
\langle R_{\rm s} \rangle = \frac{\int_0^{H_{\rm peak}} R_{\rm s}(H)S(H)H^2 dH}{\int_0^{H_{\rm peak}} S(H)H^2 dH}.
\end{equation}
Practically, $S(H)$ can be evaluated by electromagnetic simulation codes like CST or HFSS, and the integral of Eq.~(\ref{eq:david}) can be carried out numerically.
In order to obtain $R_{\rm s}$, one has to invert Eq.~(\ref{eq:david}) with a given $S(H)$ and measured $\langle R_{\rm s} \rangle$.

\begin{figure}[h]
\includegraphics[width=90mm]{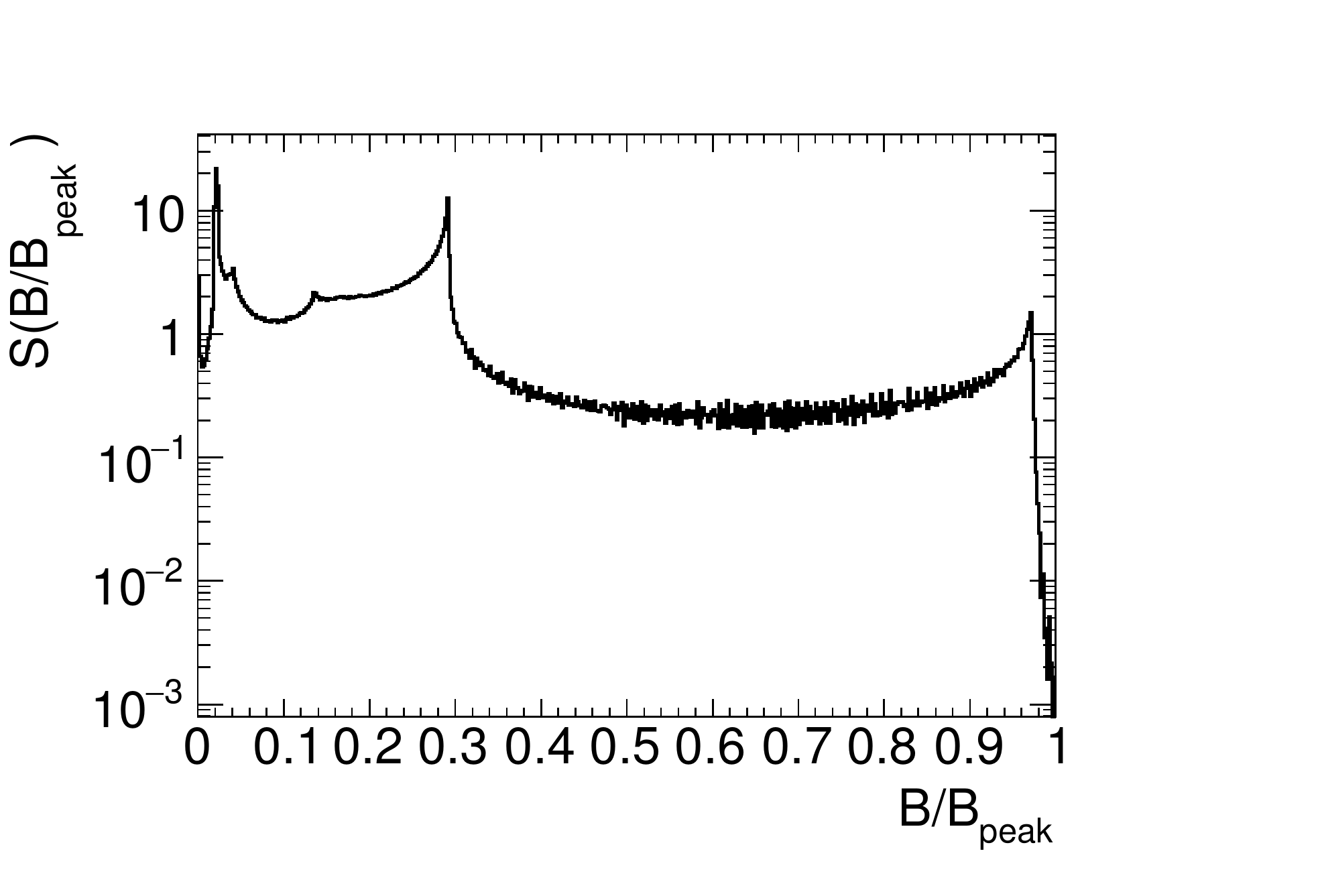}
\caption{
Area distribution function of RF field $B$ over the RF surface as a weight factor of surface integral.
\label{fig:S_vs_H}}
\end{figure}

Figure~\ref{fig:S_vs_H} shows the distribution function $S(H)$ of the HIE-ISOLDE QWR studied in this paper.
A peak in $S(H)$ means more area in corresponding RF field i.e. a local plateau of the RF field distribution.
There are two peaks in this QWR apart from the peak at the very low field corresponding to the beam holes.
One near the peak field is on the inner conductor, and the other at the medium field exists on the outer conductor.
The latter peak contributes to the surface resistance and effectively lowers the averaged $\langle R_{\rm s} \rangle$.
Hence, the measured Q-slope is underestimated.


Although aware of this effect, we did not apply the corrections for the analyses in this paper because of the following reasons.
The correction does not substantially distort the function decomposed by $R'_{\rm BCS}$, $R_{\rm fl}$ and $R_{\rm res, 0}$,
because the integral Eq.~(\ref{eq:david}) is dominated by $\delta$-function-like peaks in $S(H)$ which do not change the degree of the functions in the first order approximation.
The change in fitted $M$ defined by Eq.~(\ref{eq:def_M}) increases by about only 20\%.
We did not aim to support or exclude the Q-slope models by this precision.
Also, this correction is based on the assumption that $R_{\rm s}$ is uniform over the RF surface.
This is not very accurate in our case of a niobium film deposited on a copper substrate.
More importantly, trapped flux distribution and its oscillation should be inhomogeneous, 
which also prevents the simple application of the correction discussed here.

\section{Q-slope in welded cavities}\label{sec:Q-slope_welded}
The series production cavities for the HIE-ISOLDE project were designed as sketched in Fig.~\ref{fig:QSS} (a).
The inner conductor and outer conductor were machined separately, and welded together after shrink fitting.
The nominal penetration depth of the electron-beam welding was 2~mm, with a large spread in the production.
Thus, this location became a bottle neck for the heat conduction between inner and outer conductor.
As a result, the cavity behavior was completely different from the seamless design shown in Fig.~\ref{fig:QSS} (b) even though the coating recipe of the niobium film was the same.

The most striking behavior of these cavities is their strong dependence on cool down dynamics.
Their residual resistances show the same behavior as described in Eq.~(\ref{eq:def_Rs1_and_Rs0}) obtained for the seamless cavity with trapped flux.
Thus, it is natural to suppose that this Q-slope is also caused by trapped flux oscillation.
However, $R_{\rm s1}$ and $R_{\rm s0}$ of welded cavities are affected by the thermal gradient $\Delta T$ between top and bottom of the outer conductor when the cavities are cooled below $T_{\rm c}$.
For optimal performance, $\Delta T$ of less than 10~mK had to be achieved.
In the Linac commissioning, high $Q_0$ was obtained for most of the cavities by optimized cooling.

\begin{figure}[h]
\includegraphics[width=90mm]{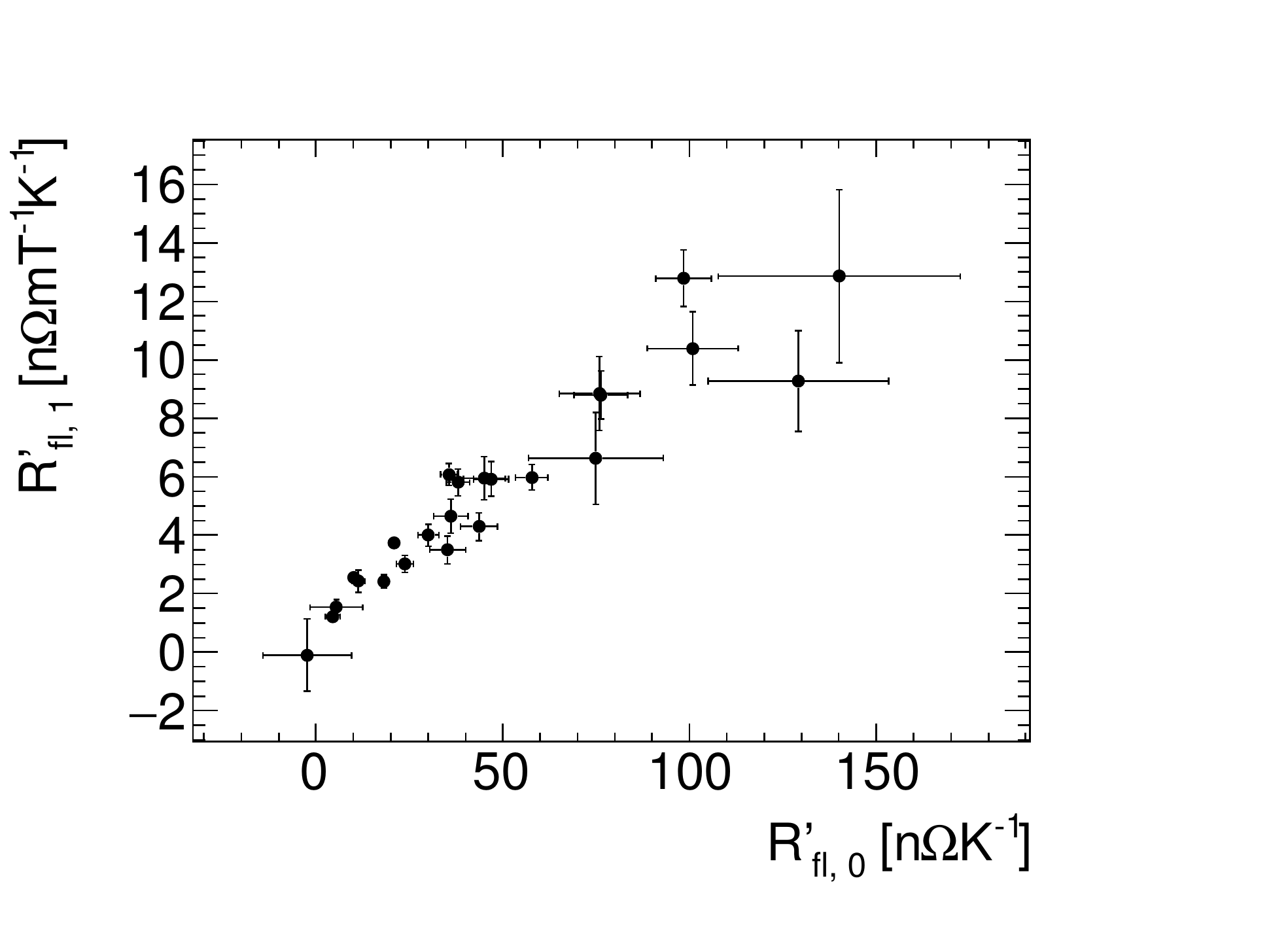}
\caption{
Correlation plot of $\Delta T$ sensitivity of $R_{\rm s1}$ and $R_{\rm s0}$.
\label{fig:Rs1DT_vs_Rs0DT}}
\end{figure}

Both the constant term $R_{\rm s0}$ and linear term $R_{\rm s1}$ depend linearly on $\Delta T$ and their sensitivities vary for each cavity.
Instead of Eq.~(\ref{eq:Rsfl_decomposition}), the flux sensitivity of the welded cavities can be expressed as
\begin{eqnarray}
R_{\rm res} & = \left[ R'_{\rm fl, 0} + R'_{\rm fl, 1} \times B_{\rm peak} \right]\times \Delta T + R_{\rm res, 0}. \label{eq:Rsfl_deltaT},
\end{eqnarray}
where $R'_{\rm fl, 0}$ and $R'_{\rm fl, 1}$ are defined as sensitivities to $\Delta T$.
A positive correlation exists between $R'_{\rm fl, 1}$ and $R'_{\rm fl, 0}$ in series production cavities as shown in Fig~\ref{fig:Rs1DT_vs_Rs0DT}.
This supports the non-linear model proposed by Ref.~\cite{vaglio18}.
Comparing Eq.~(\ref{eq:Rsfl_decomposition}) and Eq.~(\ref{eq:Rsfl_deltaT}), a natural conclusion is
\begin{equation}
H_{\rm ext} \propto \Delta T.
\end{equation}
This indicates a production of the magnetic field by thermal gradient, to be trapped by the niobium film.

A possible hypothesis is thermoelectric current inside the bimetal structure of a Nb/Cu cavity when relatively large $\Delta T$ is applied.
The required magnetic field is strong, comparable to or even higher than the earth field.
Such current may be locally confined in the interface between them:
this may have prevented its direct detection so far.

\bibliographystyle{apdrev}
\bibliography{Q-slope_ver6}
\end{document}